\documentclass[aps,prd,preprint,groupedaddress,longbibliography,nofootinbib]{revtex4-1}

\usepackage{amsmath}
\usepackage{graphicx}
\newcommand{\vev}[1]{\left\langle #1 \right\rangle}
\newcommand{\vvev}[1]{\vev{\kern-0.3em\left\langle #1
    \right\rangle\kern-0.3em}}
\newcommand{\ep}{\epsilon}
\newcommand{\ev}{\varepsilon_\textrm{vac}}
\newcommand{\Rd}[1]{\frac{\overleftarrow{\delta}}{\delta #1}}
\newcommand{\Ld}[1]{\frac{\overrightarrow{\delta}}{\delta #1}}
\newcommand{\lb}{\left\lbrace}
  \newcommand{\rb}{\right\rbrace}
\newcommand{\Tr}{\mathrm{Tr}\,}
\newcommand{\fs}[1]{\hbox{$#1$\kern-0.5em\raise0.3ex\hbox{/}}}
\newcommand{\mph}{m_\mathrm{ph}}
\newcommand{\Veff}{V_{\textrm{eff}}}

\begin{document}


\title{On the field independent additive constant in Wilson actions}


\author{Carlo Pagani} \email{cpagani@uni-mainz.de}
\affiliation{Institute f\"{u}r Physik (WA THEP)
  Johannes-Gutenberg-Universit\"{a}t\\ Staudingerweg 7, 55099 Mainz,
  Germany}

\author{Hidenori Sonoda} \email[Visiting Research Associate
till October 2024,~]{h-sonoda@pobox.com} \affiliation{Department of
  Physics and Astronomy, The University of Iowa, Iowa City, Iowa
  52242, USA}


\date{\today}

\begin{abstract}
  We discuss the field independent additive constant in Wilson actions
  carefully within the exact renormalization group formalism.  The
  additive constant does not affect the correlation functions of
  fields normalized by the partition function, and for that reason it
  is often ignored.  But it is an essential part of the partition
  function, and in the limit where the UV cutoff goes to zero, the
  constant gives a renormalized vacuum energy density.  We discuss two
  concrete examples: the Gaussian theory and the linear sigma model in
  the large $N$ limit.
\end{abstract}


\maketitle

\section{Introduction\label{Introduction}}

The Wilsonian renormalization group allows one to compute correlation
functions by performing piecemeal functional integration: rather than
integrating over all degrees of freedom at once, one lowers the
momentum cutoff gradually.  This procedure allows one to implement
several approximation schemes, which can even be of non-perturbative
nature.  If one were able to make the calculation exactly, one would
recover the correlation functions of the original microscopic model.
This holds true also for the partition function.

A particularly powerful implementation of these ideas has been
developed by Wilson \cite{Wilson:1973jj}, who put forward an exact
differential equation for the action functional associated with the
degrees of freedom that have not been integrated out yet.  Since then,
several exact renormalization group (ERG) equations have been derived
and studied
\cite{Wegner:1972ih,Polchinski:1983gv,Wetterich:1992yh,Reuter:1993kw,Morris:1993qb}.

In this paper we focus on the Wilson action and study carefully how
one can derive the partition function or equivalently the vacuum
energy density from it.
For the partition function to be preserved, the ERG must take the form
of a continuity equation, i.e., the change of the exponentiated Wilson
action under the change of the momentum cutoff $\Lambda$ must be a
total differential with respect to the fluctuating field:
\begin{eqnarray}
  -\Lambda \partial_\Lambda e^{S_\Lambda \left[\phi \right]} 
  = \int d^D x\, \frac{\delta}{\delta \phi (x)} \Bigr[ \cdots \Bigr] e^{S_\Lambda \left[\phi \right]} \,.
\end{eqnarray}
Despite its being conceptually clear from the very beginning
\cite{Wilson:1973jj,Wegner:1972ih,Wegner_1974}, in practice almost all
the works employing the Wilson action neglect the vacuum energy
density and the relevant terms in the associated ERG equation.

In the present work several aspects of the vacuum energy density are
studied and discussed in detail.  In particular, following Wilson
\cite{Wilson:1973jj}, we impose that an ERG equation have a
diffusion-inspired form mentioned above.  In so doing we shall
introduce two cutoff functions.
We highlight the role of different variables and their relation to the
limit $\Lambda \rightarrow 0$ as well as their relation to other
functionals, such as the associated effective average action.
Furthermore, we relate equivalent Wilson actions, i.e., Wilson actions
constructed via different cutoff functions whose partition functions
and correlation functions are identical \cite{Sonoda:2015bla}, by
means of a functional integral formula.

The paper is organized as follows.  In section \ref{Scalar} we
introduce our formalism for the case of a scalar field theory and
discuss the introduction of different variables and the relation
between equivalent Wilson actions.  In section \ref{Dirac} we
generalize our discussion to include fermions.  In section
\ref{sec:effective_potential} we relate the Wilson action to the
generating functional of connected correlation functions and to the
effective average action.  We show how the vacuum energy density
appears in each of these functionals.  In section \ref{Gauss} and
\ref{LargeN} we consider two examples: the Gaussian theory for scalars
and fermions and the large $N$ limit of the $\mathrm{O}\left(N\right)$
linear sigma model.  We summarize our findings in section
\ref{sec:conclusions}.  Some technical discussions are relegated to
Appendices \ref{app:lim_Lambda_to_0} and \ref{App:review_large_N}.

Throughout the paper we work on $D$-dimensional Euclidean space, and
employ the following short-hand notations:
\begin{equation}
  \int_p = \int \frac{d^D p}{(2 \pi)^D},\quad
  \delta (p) = (2 \pi)^D \delta^{(D)} (p)\,.
\end{equation}

\section{Real scalar theory\label{Scalar}}

We introduce an ultraviolet momentum cutoff $\Lambda$ in terms of a
cutoff function $R_\Lambda (p)$ that has the following properties:
\begin{enumerate}
\item It is a decreasing but positive function of $p^2$.
  \item It is an increasing function of $\Lambda$.
\item It can be written as
  \begin{equation}
    R_\Lambda (p) = \Lambda^2 R (p/\Lambda)\,,\label{RL-scalar}
  \end{equation}
  where
  \begin{subequations}
    \begin{align}
      R (0) &= 1\,,\\
      \lim_{x \to +\infty} R (x) &= 0\,.
    \end{align}
  \end{subequations}
\end{enumerate}
$R_\Lambda (p)$ gives a degree of functional integration over the
fluctuating fields of momentum $p$.  The smaller it is, the amount of
remaining integration is smaller.  Since $R_\Lambda (p)$ is a
decreasing function of $p$, the fluctuating fields of higher momenta
are integrated more than those of lower momenta.

We start with a Wilson action $S_\Lambda [\phi]$ whose modified
correlation functions \cite{Sonoda:2015bla}
\begin{equation}
  \vvev{\phi (p_1) \cdots \phi (p_n)}
  \equiv \prod_1^n \frac{1}{K (p_i/\Lambda)}\cdot
  \vev{\exp \left( - \frac{1}{2} \int_p \frac{k (p/\Lambda)}{p^2}
      \frac{\delta^2}{\delta \phi (p) \delta \phi (-p)}\right)\, \phi
    (p_1) \cdots \phi (p_n)}_{S_\Lambda}\label{mod-correlation}
\end{equation}
are independent of the cutoff $\Lambda$.  The two cutoff functions
$K(p/\Lambda)$ and $k(p/\Lambda)$ are related to $R_\Lambda (p)$ by
\begin{equation}
  R_\Lambda (p) = \frac{p^2}{k(p/\Lambda)} K(p/\Lambda)^2\,.
\end{equation}
For example, in \cite{Wilson:1973jj} the following choice is made:
\begin{equation}
  K(p/\Lambda) = \exp \left(- \frac{p^2}{\Lambda^2}\right),\quad
  k(p/\Lambda) = \frac{p^2}{\Lambda^2},\quad
  R_\Lambda (p) = \Lambda^2 \exp \left( - 2
    \frac{p^2}{\Lambda^2}\right)\,.\label{cutoff-example1}
\end{equation}
The same $R_\Lambda (p/\Lambda)$ is also obtained from
\begin{equation}
  K(p/\Lambda) = \frac{\Lambda^2 e^{- 2 \frac{p^2}{\Lambda^2}}}{p^2 +
    \Lambda^2 e^{-2\frac{p^2}{\Lambda^2}}},\quad
  k(p/\Lambda) = K(p/\Lambda)\left(1 - K(p/\Lambda)\right)
  = p^2 \frac{\Lambda^2 e^{-2\frac{p^2}{\Lambda^2}}}{\left(p^2 +
      \Lambda^2 e^{-2\frac{p^2}{\Lambda^2}}\right)^2}\,.\label{cutoff-example2}
\end{equation}

Please note that the modified correlation functions
(\ref{mod-correlation}) are not normalized by the partition function
\begin{equation}
Z_\Lambda \equiv  \int [d\phi] \exp \left( S_\Lambda [\phi] \right)\,.
\end{equation}
Hence, for $n=0$, the $\Lambda$ independence of
(\ref{mod-correlation}) amounts to the $\Lambda$ independence of the
partition function
\begin{equation}
  \frac{d}{d\Lambda} Z_\Lambda = 0\,.
\end{equation}
This implies that $\partial_\Lambda e^{S_\Lambda [\phi]}$ must be a
total differential with respect to the field variables as we explained
in Sec.~\ref{Introduction}.  Indeed we find the ERG differential
equation as \cite{Sonoda:2015bla}\footnote{We have added a constant to
  rewrite (26) of \cite{Sonoda:2015bla} as a total differential.}
\begin{equation}
  - \Lambda \partial_\Lambda e^{S_\Lambda [\phi]}
  = \int_p \frac{\delta}{\delta \phi (p)} \left[ \Lambda
    \frac{\partial \ln K(p/\Lambda)}{\partial \Lambda}\,\phi (p) + \frac{1}{2}
    \Lambda \frac{\partial R_\Lambda (p)}{\partial \Lambda} \cdot
    \frac{K(p/\Lambda)^2}{R_\Lambda (p)^2} \frac{\delta}{\delta \phi
      (-p)} \right] e^{S_\Lambda [\phi]}\,.\label{ERG-phi}
\end{equation}

We would like to replace the field variables $\phi (p)$ by alternative
field variables $\sigma (p)$ that simplify the Wilson action in the
limit $\Lambda \to 0+$.  We define
\begin{equation}
  \sigma (p) \equiv \frac{\sqrt{R_\Lambda (p)}}{K(p/\Lambda)}\, \phi
  (p) = \sqrt{\frac{p^2}{k(p/\Lambda)}}\, \phi (p)\,.\label{spin-variables}
\end{equation}
We assume that this is an analytic change of variables with respect to
the momentum $p$.  (That is the case with the two examples
(\ref{cutoff-example1},  \ref{cutoff-example2}).)  The Jacobian of this
change of variables is a field independent constant, and the Wilson
action for the $\sigma$ field is given by
\begin{equation}
  e^{S'_\Lambda [\sigma]}
  = \frac{\int [d\phi] \exp \left( - \frac{1}{2} \int_p
      \frac{p^2}{k(p/\Lambda)} \phi (p) \phi (-p)\right)}{\int
    [d\sigma] \exp \left( - \frac{1}{2} \int_p \sigma (p) \sigma
      (-p)\right)}
  \, e^{S_\Lambda [\phi]}\,.
\end{equation}
By definition, the partition function does not change:
\begin{equation}
\int [d\sigma] e^{S'_\Lambda [\sigma]}=  \int [d\phi] e^{S_\Lambda
  [\phi]}\,.
\end{equation}
In the following we adopt a particular convention of the Gaussian
functional integral:
\begin{equation}
  \int [d\sigma] \exp \left( - \frac{1}{2} \int_p \sigma (p) \sigma
    (-p) \right) = 1\,.\label{Gaussian-normalization}
\end{equation}
This implies
\begin{equation}
  \int [d\sigma] \exp \left( - \frac{1}{2} \int_p A (p) \sigma (p)
    \sigma (-p) \right) = \exp \left( - \frac{1}{2} \delta\left(0\right) \int_p \ln A (p) \right)\,,
\end{equation}
where
\begin{equation}
  \delta (0) = \int d^D x\, e^{i p \cdot 0} = V T
\end{equation}
is the volume of spacetime.

In terms of the $\sigma$ variables, the same correlation functions are
given by
\begin{equation}
  \vvev{\phi (p_1) \cdots \phi (p_n)}
  \equiv \prod_{i=1}^n \frac{1}{\sqrt{R_\Lambda (p_i)}}\cdot
  \vev{\exp \left( - \frac{1}{2} \int_p \frac{\delta^2}{\delta \sigma
        (p) \delta \sigma (-p)}\right)\,\sigma (p_1) \cdots \sigma
    (p_n)}_{S'_\Lambda}\,,
\end{equation}
The ERG equation (\ref{ERG-phi}) that guarantees the $\Lambda$
independence of the correlation functions is now given by
\begin{equation}
  - \Lambda \frac{\partial}{\partial \Lambda} e^{S'_\Lambda [\sigma]}
  = \frac{1}{2} \int_p \Lambda \frac{\partial \ln R_\Lambda
    (p)}{\partial \Lambda} \frac{\delta}{\delta \sigma (p)} \left[\left(
    \sigma (p) + \frac{\delta}{\delta \sigma (-p)} \right)
  e^{S'_\Lambda [\sigma]} \right]\,.\label{ERGeq}
\end{equation}

The Wilson action $S'_\Lambda [\sigma]$ can be interpreted so that the
field modes with $p < \Lambda$ are still to be integrated.  As we
decrease $\Lambda$ toward $0$, less and less number of degrees of
freedom are to be integrated, and eventually we obtain
\begin{equation}
  \lim_{\Lambda \to 0+} S'_\Lambda [\sigma]  = - \ev \, \delta (0) - \frac{1}{2} \int_p
  \sigma (p) \sigma (-p)\,.
  \label{eq:Ssigma_Lambda_0}
\end{equation}
We will derive this limit in Appendix \ref{app:lim_Lambda_to_0}.  Let
us note that this was already observed also by Wilson
\cite{Wilson:1973jj} in its zero dimensional version of the ERG before
discussing the full fledged ERG in terms of ``spin variables''.
The reader may be puzzled by such a simple form of
$S'_\Lambda [\sigma]$ at $\Lambda = 0$ for any microscopic model; it
seems that the physics is lost.  The crucial observation here is that
the choice of variables is very important.  As we shall see in section
\ref{sec:effective_potential}, $S'_\Lambda [\sigma]$ is related to the
generating functional of connected correlation functions.  Thus, the
physics is not lost, but in order to obtain desired results one must
adopt suitable variables before taking the limit
$\Lambda \rightarrow 0$.

Using (\ref{Gaussian-normalization}), we obtain the cutoff independent
partition function as
\begin{equation}
  \int [d\sigma] e^{S'_\Lambda [\sigma]} = \exp \left( - \ev\, \delta
    (0)\right)\,,
\end{equation}
where $\ev$ is the density of the vacuum energy.  Taking $\delta (0)$
literally, it diverges, and the partition function is either infinite
(if $\ev < 0$) or zero (if $\ev > 0$), but $\ev$ is a finite physical
quantity.  We will drop the prime from the Wilson action
$S'_\Lambda [\sigma]$ from now on.

Let us conclude this section by considering two Wilson actions
associated with the same microscopic action and built via two
different sets of cutoff functions, say $K_1,R_1$ and $K_2,R_2$.
Augmenting the result from \cite{Sonoda:2015bla} by a constant
multiple, one obtains
\begin{align}
  &e^{S_2\left[\phi\right]} =
    \frac{1}{\int \left[d\phi^{\prime\prime}\right]
    \exp \Biggr[ 
    -\frac{1}{2} \int_p \frac{1}{\frac{1}{R_2(p)}-\frac{1}{R_1(p)}}
    \frac{\phi^{\prime\prime}\left(p\right)}{K_2\left(p\right)}
    \frac{\phi^{\prime\prime}\left(-p\right)}{K_2\left(p\right)}
    \Biggr]}\notag\\
  & \times
    \int \left[d\phi^{\prime}\right]
    \exp \Biggr[
    S_1 \left[\phi^\prime \right]
    -\frac{1}{2} \int_p \frac{1}{\frac{1}{R_2(p)}-\frac{1}{R_1(p)}}
    \left(\frac{\phi^{\prime}\left(p\right)}{K_1\left(p\right)}-\frac{\phi
    \left(p\right)}{K_2\left(p\right)} \right) 
    \left(\frac{\phi^{\prime}\left(-p\right)}{K_1\left(p\right)}-\frac{\phi
    \left(-p\right)}{K_2\left(p\right)} \right) 
    \Biggr] \,. \label{eq:relation-WilsonActions} 
\end{align}
This is consistent with
\begin{eqnarray}
\int \left[d \phi \right] e^{S_1 \left[\phi \right]}
=
\int \left[d \phi \right] e^{S_2 \left[\phi \right]} \,.
\end{eqnarray}

\section{Dirac fermion theory\label{Dirac}}

It is straightforward to generalize what we have introduced for the
scalar theory to the Dirac fermion theory.  We introduce Dirac spinor
fields $\sigma (p)$ and $\bar{\sigma} (-p)$ so that the modified
correlation functions, defined by
\begin{align}
  &\vvev{\psi (p_1) \cdots \psi (p_n) \bar{\psi} (-q_1) \cdots
    \bar{\psi} (-q_n)}\notag\\
  &\equiv \prod_{i=1}^n \frac{1}{\sqrt{R_\Lambda (p_i) R_\Lambda
    (q_i)}}\, \cdot \vev{\sigma (p_1) \cdots \sigma (p_n) \exp \left(
    - \int_p \Rd{\sigma (p)} \Ld{\bar{\sigma} (-p)}\right) \bar{\sigma}
    (-q_1) \cdots \bar{\sigma} (-q_n)}_{S_\Lambda}\,,
\end{align}
are independent of $\Lambda$.  This implies the ERG equation
\begin{align}
&  - \Lambda \frac{\partial}{\partial \Lambda} e^{S_\Lambda [\sigma,
  \bar{\sigma}]}
                = - \frac{1}{2} \int_p \Lambda \partial_\Lambda \log R_\Lambda (p)\\
  &\times \Tr \left[ \Ld{\bar{\sigma} (-p)} \lb e^{S_\Lambda}
    \left(\bar{\sigma} (-p) + \Rd{\sigma (p)} \right)\rb +
    \lb \left( \sigma (p) + \Ld{\bar{\sigma} (-p)} \right)
    e^{S_\Lambda} \rb \Rd{\sigma (p)} \right]\,,
\end{align}
where the right-hand side is again a total differential.  As opposed
to (\ref{RL-scalar}) for scalar fields, we use
\begin{equation}
  R_\Lambda (p) = \Lambda R (p/\Lambda)\label{RL-Dirac}
\end{equation}
for the Dirac fermions.

In this case we find the asymptotic behavior
\begin{equation}
  \lim_{\Lambda \to 0+} S_\Lambda [\sigma, \bar{\sigma}] = -
  \ev\,\delta (0) - \int_p \bar{\sigma} (-p)  \sigma (p)\,.\label{SL-Dirac-asymptotic}
\end{equation}
Using the convention
\begin{equation}
  \int [d\sigma d\bar{\sigma}]\, \exp \left[ -  \int_p \bar{\sigma} (-p)
    \sigma (p)\right] = 1\,,
\end{equation}
we obtain
\begin{equation}
  \int [d\sigma d\bar{\sigma}] e^{S_\Lambda [\sigma, \bar{\sigma}]} =
  \exp \left( - \ev\, \delta (0)\right)\,.
\end{equation}

\section{Effective potential} \label{sec:effective_potential}

So far we have discussed how to obtain the vacuum energy density as
the limit of a Wilson action as $\Lambda \to 0+$.  In literature it is
more standard to calculate it as the minimum of the effective
potential.  Let us briefly explain the equivalence of the two
approaches.

We define
\begin{equation}
  W_\Lambda [J] \equiv S_\Lambda [\sigma] + \frac{1}{2} \int_p \sigma
  (p) \sigma (-p)\,,
\end{equation}
where
\begin{equation}
  J (p) \equiv \sqrt{R_\Lambda (p)}\, \sigma (p)\,.
\end{equation}
The ERG equation for $W_\Lambda [J]$ is obtained from that for
$S_\Lambda [\sigma]$ as
\begin{equation}
  - \Lambda \frac{\partial}{\partial \Lambda} e^{W_\Lambda [J]}
  = \frac{1}{2} \int_p \Lambda \frac{\partial R_\Lambda (p)}{\partial
    \Lambda} \frac{\delta^2}{\delta J(p) \delta J(-p)} e^{W_\Lambda
    [J]}\,.\label{ERG-W}
\end{equation}
As is explained in Appendix \ref{app:lim_Lambda_to_0}, $W_\Lambda [J]$ becomes the generating
functional of the connected correlation functions in the limit
$\Lambda \to 0+$.  Expanding $W_\Lambda [J]$ in powers of $J$, we
obtain
\begin{equation}
  W_\Lambda [J] = c_\Lambda \delta (0) + \frac{1}{2} \int_p J(p) C_{2\Lambda}
  (p) J (-p) + \cdots\,,
\end{equation}
where the constant part is the same as that in the expansion of
$S_\Lambda [\sigma]$.  The ERG equation (\ref{ERG-W}) gives
\begin{equation}
  - \Lambda \frac{\partial c_\Lambda}{\partial \Lambda}
  = \frac{1}{2} \int_p \Lambda \frac{\partial R_\Lambda (p)}{\partial
    \Lambda} \, C_{2 \Lambda} (p)\,.
\end{equation}
If we know $C_{2\Lambda} (p)$,  we can solve this to determine
$c_\Lambda$.  That is what we will do in Sec.~\ref{LargeN}.

What is usually studied in the context of the exact renormalization
group is the one-particle-irreducible (1PI) Wilson action
$\Gamma_\Lambda$, which is defined as the Legendre transform of
$W_\Lambda [J]$:
\begin{equation}
  \Gamma_\Lambda [\Phi] - \frac{1}{2} \int_p R_\Lambda (p) \Phi (p)
  \Phi (-p) \equiv W_\Lambda [J] - \int_p J(-p) \Phi (p)\,,\label{Gamma-W}
\end{equation}
where
\begin{equation}
  \Phi (p) \equiv \frac{\delta W_\Lambda [J]}{\delta J (-p)}\,.
\end{equation}
In the limit $\Lambda \to 0+$, we obtain the effective action
\begin{equation}
  \lim_{\Lambda \to 0+} \Gamma_\Lambda [\Phi] = \Gamma_\textrm{eff}
  [\Phi]\,.
\end{equation}

For constant fields $J (p) = j\, \delta (p)$ and $\Phi (p) = \varphi\,\delta (p)$, we obtain
\begin{equation}
W_\Lambda [J] = w_\Lambda (j)\,\delta (0),\quad  \Gamma_\Lambda [\Phi]
= G_\Lambda (\varphi)\, \delta (0)\,. 
\end{equation}
Eq.~(\ref{Gamma-W}) reduces to the Legendre transformation
\begin{equation}
  G_\Lambda (\varphi) - \frac{1}{2} \Lambda^2 \varphi^2 = w_\Lambda
  (j) - j \varphi\,,
\end{equation}
where
\begin{equation}
  \varphi = w'_\Lambda (j)\,.
\end{equation}
The inverse Legendre transformation gives
\begin{equation}
   j = - G'_\Lambda (\varphi)\,.
\end{equation}
We assume that $j=0$ corresponds to the minimum, not the maximum, of
$- G_\Lambda (\varphi)$.  The effective potential is the limit
\begin{equation}
  \Veff (\varphi) = - \lim_{\Lambda \to 0+} G_\Lambda (\varphi)\,.
\end{equation}

Let $v_\Lambda$ be the value of $\varphi$ at the minimum of
$- G_\Lambda (\varphi)$, satisfying
\begin{equation}
  G'_\Lambda (v_\Lambda) = 0\,.
\end{equation}
We then obtain
\begin{equation}
 c_\Lambda = w_\Lambda (0) = G_\Lambda (v_\Lambda) - \frac{1}{2}   \Lambda^2 v_\Lambda^2\,.
\label{eq:cLambda_via_G_Lambda_vL}
\end{equation}
Hence, we obtain
\begin{equation}
  \ev = - \lim_{\Lambda \to 0+} c_\Lambda = - \lim_{\Lambda \to 0+}
  G_\Lambda (v_\Lambda) = \Veff (v)\,,
\end{equation}
where $v$ is at the minimum of the effective potential satisfying
\begin{equation}
  \Veff' (v) = 0\,.
\end{equation}

\section{The Gaussian theory\label{Gauss}}

In this section we compute $\ev$ for the Gaussian theory, both bosonic
and fermionic.

\subsection{Scalar theory\label{scalar}}

We assume a quadratic form
\begin{equation}
  S_\Lambda [\sigma] = c_\Lambda\, \delta (0) - \frac{1}{2} \int_p
  \sigma (p) C_\Lambda (p) \sigma (-p)\,.
\end{equation}
Substituting this into (\ref{ERGeq}), we obtain
\begin{subequations}
  \begin{align}
    - \Lambda \frac{\partial}{\partial \Lambda} c_\Lambda
    &= \frac{1}{2} \int_p \Lambda \frac{\partial \ln R_\Lambda
      (p)}{\partial \Lambda} \, \left(1 - C_\Lambda (p)\right)\,,\label{cLambda-Gauss}\\
    \Lambda \frac{\partial}{\partial \Lambda} C_\Lambda (p)
    &= - \Lambda \frac{\partial \ln R_\Lambda (p)}{\partial \Lambda}
      \left(1-C_\Lambda (p)\right) C_\Lambda (p)\,.
  \end{align}
\end{subequations}
The second equation can be solved as
\begin{equation}
  C_\Lambda (p) = \frac{F (p)}{F (p) + R_\Lambda (p)}\,,
\end{equation}
where $F (p)$ is independent of $\Lambda$.  This corresponds to the
correlation function
\begin{equation}
  \vvev{\phi (p) \phi (q)} = \delta (p+q)\, \frac{1}{F (p)} \times
  e^{- \ev\,\delta (0)}\,.
\end{equation}
The Gaussian theory is given by
\begin{equation}
  F (p) = p^2 + m^2\,,
\end{equation}
where $m^2$ is a constant squared mass.

Eq.~(\ref{cLambda-Gauss}) now gives
\begin{equation}
  - \Lambda \frac{\partial}{\partial \Lambda} c_\Lambda = \frac{1}{2} \int_p \Lambda
  \frac{\partial R_\Lambda (p)}{\partial \Lambda} \, \frac{1}{p^2 +
    m^2 + R_\Lambda (p)}\,.\label{cLambda-scalar}
\end{equation}
We will solve this first for $2 < D < 4$, and then for $D=4$.

For the scalar theory, $R_\Lambda (p) = \Lambda^2 R (p/\Lambda)$, and
\begin{equation}
  \Lambda \frac{\partial R_\Lambda (p)}{\partial \Lambda}
  = \left( 2 - p \cdot \partial_p \right) R_\Lambda (p)\,.
\end{equation}
For $2 < D < 4$, we can rewrite (\ref{cLambda-scalar}) as
\begin{align}
  - \Lambda \frac{\partial c_\Lambda}{\partial \Lambda}
  &= \Lambda^D \int_p \left(1 - \frac{1}{2} p \cdot \partial_p \right)
    R (p) \cdot \frac{1}{p^2 + \frac{m^2}{\Lambda^2} + R (p)}\notag\\
  &= \Lambda^D \int_p \left(1 - \frac{1}{2} p \cdot \partial_p \right)
    R (p) \cdot  \left[ \frac{1}{p^2 + R(p)} - \frac{m^2}{\Lambda^2}
    \frac{1}{\left( p^2 + R (p)\right)^2}\right]\notag\\
  &\quad + \frac{1}{2} \int_p \Lambda \frac{\partial R_\Lambda
    (p)}{\partial \Lambda} \left[ \frac{1}{p^2 + m^2 + R_\Lambda (p)} -
    \frac{1}{p^2 + R_\Lambda (p)} + \frac{m^2}{\left(p^2 + R_\Lambda
    (p)\right)^2} \right]\,.
\end{align}
Integrating this, we obtain
\begin{align}
  c_\Lambda
  &= c - \frac{1}{D} \Lambda^D \int_p \left(1 - \frac{1}{2} p \cdot
    \partial_p \right) R (p) \cdot \frac{1}{p^2+R(p)}\notag\\
  &\quad + \frac{1}{D-2} m^2 \Lambda^{D-2} \int_p \left(1 - \frac{1}{2} p \cdot
    \partial_p \right) R (p) \cdot
    \frac{1}{\left(p^2+R(p)\right)^2}\notag\\
  &\quad - \frac{1}{2} \int_p \left[ \ln \frac{p^2 + m^2 + R_\Lambda
    (p)}{p^2 + R_\Lambda (p)} - \frac{m^2}{p^2 + R_\Lambda (p)}
    \right]\,,
\end{align}
where $c$ is a constant of integration with mass dimension $D$, and
the last integral is UV finite.  Hence, we obtain
\begin{align}
  \ev
  &= - \lim_{\Lambda \to 0+} c_\Lambda\notag\\
  &= - c + \frac{1}{2} \int_p \left[ \ln \frac{p^2+m^2}{p^2} -
    \frac{m^2}{p^2} \right]\,,\label{ev-Gauss-scalar}
\end{align}
where the integral is UV finite, and we obtain
\begin{align}
  \frac{1}{2} \int_p \left[ \ln \frac{p^2+m^2}{p^2} -
  \frac{m^2}{p^2} \right]
  &= \frac{m^2}{D} \int_p \left[ 
    \frac{1}{p^2 + m^2} - \frac{1}{p^2} \right]\notag\\
  &= - \frac{1}{2} \frac{1}{(4 \pi)^{\frac{D}{2}}} \Gamma \left( -
    \frac{D}{2}\right)\, (m^2)^{\frac{D}{2}} \,,\label{result-Gaussian-scalar}
\end{align}
which is negative for $2 < D < 4$.  $\ev$ is not analytic with respect
to $m^2$.  This is equal to the zero-point energy calculated with
dimensional regularization:
\begin{equation}
  \frac{1}{2} \int \frac{d^{D-1} p}{(2 \pi)^{D-1}} \sqrt{\vec{p}\,^2 +
    m^2}
  = - \frac{1}{2} \frac{1}{(4 \pi)^{\frac{D}{2}}} \Gamma \left( -
    \frac{D}{2}\right)\, (m^2)^{\frac{D}{2}}\,.
\end{equation}

For $D=4$, we can rewrite (\ref{cLambda-scalar}) as
\begin{align}
&  - \Lambda \frac{\partial c_\Lambda}{\partial \Lambda}
  = \int_p \left(1 - \frac{1}{2} p \cdot \partial_p \right)
    R (p) \cdot  \left[ \Lambda^4 \frac{1}{p^2 + R(p)} - m^2 \Lambda^2
    \frac{1}{\left( p^2 + R (p)\right)^2} + m^4 \frac{1}{\left(p^2 +
   R (p)\right)^3}\right]\notag\\
  &\quad + \frac{1}{2} \int_p \Lambda \frac{\partial R_\Lambda
    (p)}{\partial \Lambda} \left[ \frac{1}{p^2 + m^2 + R_\Lambda (p)} -
    \frac{1}{p^2 + R_\Lambda (p)} + \frac{m^2}{\left(p^2 + R_\Lambda
    (p)\right)^2} - \frac{m^4}{\left(p^2 + R_\Lambda
    (p)\right)^3}\right]\,,
    \label{dcLambda-4}
\end{align}
where the integral multiplied by $m^4$ is obtained as
\begin{align}
 \int_p \left(1 - \frac{1}{2} p \cdot \partial_p \right) R (p) \cdot
  \frac{1}{\left(p^2 + R(p)\right)^3}
  &= \frac{1}{4} \int_p \left(p
    \cdot \partial_p + 4 \right) \frac{1}{\left(p^2 + R
    (p)\right)^2}\notag\\
  &=\frac{1}{4} \frac{2}{(4 \pi)^2} \int_0^\infty dp^2 \frac{d}{dp^2}
    \frac{p^4}{\left(p^2 + R (p)\right)^2} = \frac{1}{2 (4 \pi)^2}\,.
\end{align}
Integrating (\ref{dcLambda-4}) over $\Lambda$, we obtain
\begin{align}
  c_\Lambda
  &= c - \frac{\Lambda^4}{4} \int_p \left(1 - \frac{1}{2} p \cdot
    \partial_p \right) R (p) \cdot \frac{1}{p^2 + R (p)} + \frac{m^2
    \Lambda^2}{2} \int_p \left(1 - \frac{1}{2} p \cdot
    \partial_p \right) R (p) \cdot \frac{1}{\left(p^2 + R
      (p)\right)^2}\notag\\
  &\quad - \frac{1}{2 (4 \pi)^2} m^4 \ln \frac{\Lambda}{\mu} +
    \Lambda^4 F \left(\frac{m^2}{\Lambda^2}\right)\,,
\end{align}
where $c$ is a constant of integration, $\mu$ is an arbitrary mass
parameter, and
\begin{align}
  \Lambda^4 F \left(\frac{m^2}{\Lambda^2}\right)
  &\equiv \frac{1}{2} \int_\Lambda^\infty d\Lambda'
    \frac{\partial}{\partial \Lambda'} \int_p \left[
 \ln \frac{p^2 + R_{\Lambda'} (p)+m^2}{p^2+R_{\Lambda'} (p)} -
    \frac{m^2}{p^2 + R_{\Lambda'} (p)} + \frac{1}{2}\frac{m^4}{\left(p^2 + R_{\Lambda'}
        (p)\right)^2} \right]\notag\\
    &=- \frac{1}{2} \int_p \left[
    \ln \frac{p^2 + R_\Lambda (p)+m^2}{p^2+R_\Lambda (p)} -
    \frac{m^2}{p^2 + R_\Lambda (p)} + \frac{1}{2} \frac{m^4}{\left(p^2 + R_\Lambda
        (p)\right)^2} \right]\,.
\end{align}
By definition we find
\begin{equation}
\Lambda^4 F \left(\frac{m^2}{\Lambda^2}\right) \overset{\Lambda \to
  \infty}{\longrightarrow} 0\,.
\end{equation}
Since $c_\Lambda$ remains finite as $\Lambda \to 0+$, we obtain
\begin{equation}
  \Lambda^4 F \left(\frac{m^2}{\Lambda^2}\right) \overset{\Lambda \to
  0+}{\longrightarrow}  - \frac{1}{4 (4 \pi)^2} m^4 \ln
\frac{m^2}{\Lambda^2} + \mathrm{const} \times m^4\,.
\end{equation}
Hence, we obtain
\begin{equation}
  \ev = - \lim_{\Lambda \to 0+} c_\Lambda = - c + \frac{1}{4 (4
    \pi)^2} m^4 \ln \frac{m^2}{\mu^2}\,,
    \label{eq:vac_energy_density_free_scalar}
\end{equation}
where we have redefined $c$ by absorbing a constant multiple of $m^4$.
The change of $\mu$ can be compensated by a change of $c$.  This
result is consistent with the zero-point energy obtained by
dimensional regularization with a minimal subtraction ($D = 4 - \ep$):
\begin{equation}
  \mu^\ep \frac{1}{2} \int \frac{d^{D-1} p}{(2 \pi)^{D-1}} \sqrt{\vec{p}\,^2 +
    m^2} + \frac{1}{\ep} \frac{m^4}{2 (4 \pi)^2} 
  \overset{\ep \to 0}{\longrightarrow} \frac{m^4}{4 (4 \pi)^2} \ln
  \frac{m^2 e^{\gamma - \frac{3}{2}}}{4 \pi \mu^2}\,.
\end{equation}
  
Let us conclude this section by commenting on the massless case which
would be the Gaussian fixed point in the dimensionless framework.
From Eqs.~(\ref{ev-Gauss-scalar}) and
(\ref{eq:vac_energy_density_free_scalar}), we expect that the vacuum
energy density vanishes for the fixed point theory.  We find it
instructive to see how this happens in some details.  In the
dimensionless framework, where all the physical quantities are
rendered dimensionless by using appropriate powers of the cutoff
$\Lambda$, the Gaussian theory is given by
\begin{equation}
  \bar{S}_t \left[ \bar{\sigma}\right] = \bar{c}_t \,\delta (0) -
  \frac{1}{2} \int_p \bar{\sigma} (-p) \frac{p^2 + \bar{m}_t^2 }{p^2 +
    \bar{m}_t^2 + R (p)} \bar{\sigma} (p)\,,
\end{equation}
where the logarithmic scale $t$ is introduced by $\Lambda = \mu
e^{-t}$, and
\begin{equation}
  \bar{c}_t \equiv \frac{c_\Lambda}{\Lambda^D},\quad
  \bar{m}_t^2 \equiv \frac{m^2}{\Lambda^2} = \frac{m^2}{\mu^2}
  e^{2t}\,.
\end{equation}
Eq.~(\ref{cLambda-scalar}) gives
\begin{eqnarray}
\left(\frac{d}{dt}-D\right) \bar{c}_t 
=
\int_p \left(1-\frac{1}{2}p\partial_p \right) R \left(p\right)
\frac{1}{p^2+\bar{m}^2_t+R\left(p\right)}\,.
\end{eqnarray}
At the fixed-point $\bar{m}_t^2 = 0$, we obtain
\begin{equation}
\bar{S}^* \left[\bar{\sigma}\right] =
\bar{c}^* \delta \left(0\right) 
-\frac{1}{2} \int_p \bar{\sigma}\left(-p\right)
\frac{p^2}{p^2+R\left(p\right)}
 \bar{\sigma}\left(p\right)\,,
 \label{eq:GFP_S}
\end{equation}
where the fixed point value $\bar{c}^*$ is given by\footnote{ This
  result is reminiscent of the fixed point value for the cosmological
  constant.  See, for example,
  \cite{Reuter:1996cp,Reuter_Saueressig_2019,Percacci_book_2017}.}
\begin{eqnarray}
\bar{c}^* =
-\frac{1}{D} \int_p \left(1-\frac{1}{2}p\partial_p \right) R \left(p\right)
\frac{1}{p^2+ R\left(p\right)}\,.
\end{eqnarray}
By integrating over the fluctuating field $\bar{\sigma}$ in
(\ref{eq:GFP_S}), we obtain the vanishing vacuum energy density
\begin{equation}
  \bar{\varepsilon}^* =
  -\bar{c}^*+\frac{1}{2}\int_p \log \frac{p^2}{p^2+R\left(p\right)} = 0\,.
\end{equation}
Though this result is expected, it is still nice to get it by
integrating out the fluctuating field.

\subsection{Dirac theory\label{dirac}}

We assume a quadratic form
\begin{equation}
  S_\Lambda [\sigma, \bar{\sigma}] = c_{F, \Lambda} \delta (0) - \int_p
  \bar{\sigma} (-p) C_{F, \Lambda} (p) \sigma (p)\,.
\end{equation}
We obtain
\begin{equation}
  C_{F, \Lambda} (p) = \frac{i \left(\fs{p} + i m\right)}{R_\Lambda
    (p) + i \left(\fs{p} + i m\right)}\,,
\end{equation}
where
\begin{equation}
  R_\Lambda (p) = \Lambda R (p/\Lambda)\,.
\end{equation}
This corresponds to
\begin{equation}
  \vvev{\psi (p) \bar{\psi} (-q)} = \delta (p-q) \cdot \frac{1}{i
    \left(\fs{p} + i m\right)}\, e^{- \ev\,\delta (0)}\,.
\end{equation}

The ERG equation for $c_{F, \Lambda}$ is given by
\begin{align}
  \Lambda \frac{\partial c_{F, \Lambda}}{\partial \Lambda}
  &= \int_p  \Lambda \partial_\Lambda R_\Lambda (p) \cdot \Tr \frac{1}{R_\Lambda
    (p) + i \left(\fs{p} + i m\right)}\\
  &= \Tr 1 \cdot \int_p  \Lambda \partial_\Lambda R_\Lambda (p) \cdot
    \frac{R_\Lambda (p) - m}{\left(R_\Lambda (p)-m\right)^2 + p^2}\,.
\end{align}
The solution to this equation is analogous to the scalar case.  We
give it only for $3 < D < 4$:
\begin{align}
  c_{F, \Lambda}
  &= c_F +  \Tr 1 \cdot \Bigg[ \frac{\Lambda^D}{D} \int_p (1 - p \cdot \partial_p) R (p) \cdot
            \frac{R(p)}{p^2 + R(p)^2}\notag\\
  &\quad + \frac{m \Lambda^{D-1}}{D-1}  \int_p (1 - p \cdot \partial_p) R (p) \cdot
            \frac{R(p)^2 - p^2}{\left(p^2 + R(p)^2\right)^2}\notag\\
  &\quad + m^2 \frac{\Lambda^{D-2}}{D-2} \int_p  (1 - p
    \cdot \partial_p) R (p) \cdot 
    \frac{R(p)^3 - 3 p^2 R(p)}{\left(p^2 + R(p)^2\right)^3}\notag\\
  &\quad +  m^3 \frac{\Lambda^{D-3}}{D-3} \int_p
    (1-p\cdot \partial_p ) R (p) \cdot \frac{p^4 - 6 p^2 R(p)^2 +
    R(p)^4}{(p^2 + R(p)^2)^4}\notag\\
          &\quad + \int_p \lb \frac{1}{2}
    \ln \frac{\left(R_{\Lambda} (p) - m\right)^2 + p^2}{p^2 +
    R_{\Lambda} (p)^2} + m \frac{R_{\Lambda} (p)}{p^2 + R_{\Lambda}
    (p)^2} + m^2 \frac{- p^2 + R_{\Lambda} (p)^2}{2 \left(p^2 +
            R_{\Lambda} (p)^2\right)^2}\right.\notag\\
  &\left.\qquad + m^3 \frac{R_\Lambda (p) \left(R_\Lambda (p)^2-3
    p^2\right)}{3 \left(p^2 + R_\Lambda (p)^2\right)^3} \rb \Bigg]\,,
\end{align}
where $c_F$ is a constant of integration.  Taking the limit
$\Lambda \to 0+$, we obtain
\begin{align}
  \ev
  &= - c_F - \Tr 1 \cdot \frac{1}{2}  \int_p \left( \ln
    \frac{p^2+m^2}{p^2} - \frac{m^2}{p^2} \right)\notag\\
  &= - c_F + \Tr 1 \cdot \frac{1}{2} \frac{1}{(4 \pi)^{\frac{D}{2}}}
    \Gamma \left(- \frac{D}{2}\right) (m^2)^{\frac{D}{2}}\,.
\end{align}
This is consistent with the energy density of the Dirac vacuum:
\begin{equation}
-  \frac{1}{2} \Tr 1 \cdot \int \frac{d^{D-1} p}{(2 \pi)^{D-1}}
  \sqrt{\vec{p}\,^2 + m^2}\,.
\end{equation}
The factor $\frac{1}{2}$ is there because only the negative energy
states contribute to the vacuum.

\section{The large $N$ limit\label{LargeN}}

We consider the $\mathrm{O}(N)$ linear sigma model in $D$ dimensions ($2 < D < 4$).  Let
$S_\Lambda [\sigma]$ be the Wilson action in terms of the spin
variables $\sigma^I (p)\,(I=1,\cdots,N)$.  Expanding $S_\Lambda
[\sigma]$ in powers of fields, we obtain
\begin{equation}
  S_\Lambda [\sigma] = c_\Lambda\, \delta (0) + \frac{1}{2} \int_p
  \sigma^I (p) \sigma^I (-p)\, c_{2 \Lambda} (p) + \cdots\,,
\end{equation}
where the repeated indices are summed.  The ERG equation
\begin{equation}
  - \Lambda \frac{\partial}{\partial \Lambda} e^{S_\Lambda [\sigma]}
  = \frac{1}{2} \int_p \Lambda \frac{\partial \log R_\Lambda (p)}{\partial
    \Lambda} \frac{\delta}{\delta \sigma^I (p)} \left[
    \left(\sigma^I (p) + \frac{\delta}{\delta \sigma^I (-p)} \right)
    e^{S_\Lambda [\sigma]} \right]
\end{equation}
gives
\begin{equation}
  - \Lambda \frac{\partial}{\partial \Lambda} c_\Lambda
  = \frac{1}{2} \int_p \Lambda \frac{\partial \ln R_\Lambda (p)}{\partial
    \Lambda} N \left(1 + c_{2\Lambda} (p)\right)\,.\label{cL-diffeq-largeN}
\end{equation}
In the large $N$ limit, we find
\begin{equation}
  1 +  c_{2\Lambda} (p) = \frac{R_\Lambda (p)}{p^2 + R_\Lambda (p) +
    m_\Lambda^2}\,,
  \label{CL-largeN}
\end{equation}
where the squared mass $m_\Lambda^2$ satisfies
\begin{equation}
- \Lambda \frac{\partial m_\Lambda^2}{\partial \Lambda} =
\frac{\frac{1}{2} \int_p \Lambda \partial_\Lambda R_\Lambda (p)
  \frac{1}{\left(p^2 + R_\Lambda (p) +
      m_\Lambda^2\right)^2}}{\frac{1}{\lambda} + \frac{1}{2} \int_p
  \frac{1}{\left(p^2 + R_\Lambda (p) + m_\Lambda^2\right)^2}}\,.
\label{eq:flow_m2L_large_N}
\end{equation}
The positive constant $\lambda$ is a $\phi^4$ self-coupling.
Eq.~(\ref{cL-diffeq-largeN}), combined with (\ref{CL-largeN}), is the
same as Eq.~(\ref{cLambda-scalar}) for the Gaussian theory, except for
the cutoff dependence of the squared mass.  We refer the reader to
Appendix \ref{App:review_large_N} for a derivation of
Eqs.~(\ref{cL-diffeq-largeN}-\ref{eq:flow_m2L_large_N}).

The solution to (\ref{cL-diffeq-largeN}) is given by
\begin{equation}
\frac{1}{N} c_\Lambda = c - \frac{1}{2 \lambda} m_\Lambda^4 + m_\Lambda^2
\frac{1}{2} \int_p \left(\frac{1}{p^2+R_\Lambda (p) + m_\Lambda^2} -
  \frac{1}{p^2}\right)  - \frac{1}{2} \int_p \left( \ln \frac{p^2 +
    R_\Lambda (p) + m_\Lambda^2}{p^2} - \frac{m_\Lambda^2}{p^2}\right)\,,
\end{equation}
where $c$ is a constant of integration.  Denoting the physical squared
mass by
\begin{equation}
  \mph^2 = \lim_{\Lambda \to 0+} m_\Lambda^2\,,
\end{equation}
we obtain the vacuum energy density as
\begin{align}
\frac{1}{N}  \ev
  &= - \lim_{\Lambda \to 0+} \frac{1}{N} c_\Lambda\notag\\
  &= - c + \frac{1}{2 \lambda} \mph^4 - \mph^2 \frac{1}{2} \int_p
    \left(\frac{1}{p^2 + \mph^2} - \frac{1}{p^2}\right) + \frac{1}{2}
    \int_p \left( \ln \frac{p^2 + \mph^2}{p^2} -
    \frac{\mph^2}{p^2}\right)\notag\\
  &= - c + \frac{1}{2 \lambda} \mph^4 + \frac{D-2}{4} \frac{\Gamma
    \left(- \frac{D}{2}\right)}{(4 \pi)^{\frac{D}{2}}}
    (\mph^2)^{\frac{D}{2}}\,,\label{ev-largeN}
\end{align}
where $2 < D < 4$.  With $c$ and $\mph^2$ fixed, this has a well defined
strong coupling limit
\begin{equation}
  \frac{1}{N}  \ev \overset{\lambda \to \infty}{\longrightarrow} c +
  \frac{D-2}{4} \frac{\Gamma  \left(- \frac{D}{2}\right)}{(4
    \pi)^{\frac{D}{2}}} (\mph^2)^{\frac{D}{2}}\,.
\end{equation}

For (\ref{ev-largeN}) to have a good weak coupling limit, we must
introduce appropriate $\lambda$ dependence to $c$.  We note
\begin{equation}
k \equiv  - \frac{m_\Lambda^2}{\lambda} + \frac{1}{2} \int_p \left( \frac{1}{p^2 +
      R_\Lambda (p) + m_\Lambda^2} - \frac{1}{p^2}\right)
\end{equation}
is independent of $\Lambda$.  Writing
\begin{equation}
  c = c' + \frac{\lambda}{2} k^2\,,
\end{equation}
we obtain an alternative expression for
$c_\Lambda$:
\begin{equation}
\frac{1}{N} c_\Lambda =  c' + \frac{\lambda}{2} \lb\frac{1}{2} \int_p
    \left(\frac{1}{p^2 + R_\Lambda (p) + m_\Lambda^2} - \frac{1}{p^2}
              \right)\rb^2
 - \frac{1}{2} \int_p \left( \ln \frac{p^2 + R_\Lambda (p) +
    m_\Lambda^2}{p^2} - \frac{m_\Lambda^2}{p^2} \right)\,.
\end{equation}
This gives
\begin{equation}
  \frac{1}{N} \ev = - c' - \frac{\lambda}{2} 
\left( \frac{1}{\left(4\pi \right)^{D/2}} \frac{D}{4}
    \Gamma \left(-\frac{D}{2}\right) (\mph^2)^{\frac{D-2}{2}} \right)^2
    - \frac{1}{2} \frac{\Gamma
    \left(-\frac{D}{2}\right)}{(4 \pi)^{\frac{D}{2}}}
  (\mph^2)^{\frac{D}{2}} \,.
\end{equation}
In the weak coupling limit, where $c'$ is fixed instead of $c$, we
reproduce the Gaussian result (\ref{result-Gaussian-scalar}):
\begin{equation}
\frac{1}{N} \ev \overset{\lambda \to 0}{\longrightarrow} - c' - \frac{1}{2} \frac{\Gamma
    \left(-\frac{D}{2}\right)}{(4 \pi)^{\frac{D}{2}}}
    (\mph^2)^{\frac{D}{2}} \,.
\end{equation}

\section{Conclusions} \label{sec:conclusions}

In this work we have studied the field independent part of the Wilson
action that is fundamental in the calculation of the partition
function.
Let us summarize our results.

We have considered a general ERG equation for the Wilson action
involving two cutoff functions.  This equation has a diffusion-like
form for both scalar fields and fermions.
We have introduced the field variables $\sigma$ that simplify the form
of the Wilson action in the limit $\Lambda \rightarrow 0$, as given by
(\ref{eq:Ssigma_Lambda_0}) and (\ref{SL-Dirac-asymptotic}).  Moreover,
in Eq.~(\ref{eq:relation-WilsonActions}), we have formally related two
equivalent Wilson actions, constructed by two different sets of cutoff
functions, that give rise to the same correlation functions and
partition function.  We have also related our formalism to the
generating functional for connected correlation functions and to the
1PI generating functional in Sec.~\ref{sec:effective_potential}.
Finally, we have made our discussion concrete by considering two
examples: the Gaussian models for scalar and fermionic fields
(Sec.~\ref{Gauss}) and the large $N$ limit of the
$\mathrm{O}\left(N\right)$ linear sigma model (Sec.~\ref{LargeN}).
We hope this work lays a ground for further generalizations.  For
instance, it may be natural to go beyond the flat space and work on
more general background spacetimes.

\appendix
\section{The limit of $S_\Lambda [\sigma]$ as $\Lambda \to 0+$} \label{app:lim_Lambda_to_0}

In the limit $\Lambda \to 0+$ we expect $S_\Lambda [\sigma]$ to become
``trivial'' since below the UV cutoff $\Lambda = 0$ there is no mode left
to be integrated.  To make this statement precise let us compute
\[
  \lim_{\Lambda \to 0+} S_\Lambda [\sigma]
\]
explicitly, where $\sigma$ is a special choice of field variables
defined by (\ref{spin-variables}).

For this goal, we take a rather roundabout path.  We first consider
the generating functional $\mathcal{W} [\mathcal{J}]$ of the connected
correlation functions defined by \cite{Sonoda:2017rro}
\begin{align}
  e^{\mathcal{W} [\mathcal{J}]}
  &\equiv \sum_0^\infty \frac{1}{n!} \int_{p_1, \cdots, p_n}
    \mathcal{J} (-p_1) \cdots \mathcal{J} (-p_n)\, \vvev{\phi (p_1)
    \cdots \phi (p_n)}\\
  &= \sum_0^\infty \frac{1}{n!} \int_{p_1, \cdots, p_n}
    \prod_1^n \frac{\mathcal{J} (-p_i)}{\sqrt{R_\Lambda (p_i)}}
\cdot \vev{\exp \left( - \frac{1}{2} \int_p
    \frac{\delta^2}{\delta \sigma (p) \delta \sigma (-p)}\right) \sigma (p_1) \cdots \sigma
    (p_n)}_{S_\Lambda}\notag\\
  &= \int [d\sigma] e^{S_\Lambda [\sigma]} \exp \left( - \frac{1}{2} \int_p
    \frac{\delta^2}{\delta \sigma (p) \delta \sigma (-p)}\right) \exp
    \left(\int_p \frac{\mathcal{J} (-p)}{\sqrt{R_\Lambda (p)}} \sigma (p) \right)\notag\\
  &= \int [d\sigma] \exp \left[ S_\Lambda [\sigma] - \frac{1}{2} \int_p
    \frac{\mathcal{J} (p) \mathcal{J}
    (-p)}{R_\Lambda (p)} + \int_p \frac{\mathcal{J}
    (-p)}{\sqrt{R_\Lambda (p)}} \sigma (p) \right]\,.
\end{align}
We now introduce new field variables by
\begin{equation}
  J (p) \equiv   \sqrt{R_\Lambda (p)}\, \sigma (p)\,.
\end{equation}
Taking the Jacobian into account, we obtain
\begin{align}
  e^{\mathcal{W} [\mathcal{J}]}
 &= \frac{1}{\int [dJ'] \exp \left(- \frac{1}{2} \int_p \frac{J' (p)
   J'(-p)}{R_\Lambda (p)}\right)}
 \int [dJ] \exp \Bigg[ W_\Lambda [J] \notag\\
  &\qquad - \frac{1}{2} \int_p \frac{1}{R_\Lambda (p)}
   \left(\mathcal{J} (p) - J(p)\right)\left(\mathcal{J} (-p) -
     J(-p)\right)\Bigg]\,,\label{Wcal}
\end{align}
where
\begin{equation}
  W_\Lambda [J] \equiv S_\Lambda [\sigma]
  + \frac{1}{2} \int_p \sigma (p) \sigma (-p)
  = S_\Lambda \left[ \frac{J}{\sqrt{R_\Lambda}}\right]
  + \frac{1}{2} \int_p \frac{J (p) J(-p)}{R_\Lambda (p)}\,,\label{Appendix-WL}
\end{equation}
and we have used the triviality of the Gaussian functional integral
\begin{equation}
  \int [d\sigma]\,\exp \left( - \frac{1}{2} \int_p \sigma (p) \sigma
    (-p)\right) = 1\,.
\end{equation}
The right-hand side of (\ref{Wcal}) is independent of $\Lambda$.  In
the limit $\Lambda \to 0+$ the functional integral becomes one over
the delta functional
\begin{align}
&  \lim_{\Lambda \to 0+}
  \frac{1}{\int [dJ'] \exp \left(-\frac{1}{2} \int_p \frac{J'(p)
        J'(-p)}{R_\Lambda (p)}\right)}
  \exp \left( - \frac{1}{2} \int_p \frac{1}{R_\Lambda (p)}
    \left(\mathcal{J} (p) - J(p)\right)\left(\mathcal{J} (-p) -
      J(-p)\right)\right)\notag\\
&= \prod_p \delta \left( \mathcal{J} (p) - J (p)\right)\,,
\end{align}
and we obtain
\begin{equation}
  \lim_{\Lambda \to 0+} W_\Lambda [J] = \mathcal{W} [J]\,.
\end{equation}
Fixing $\sigma (p)$, we obtain
\begin{equation}
  \lim_{\Lambda \to 0+} J (p) = 0\,.
\end{equation}
Hence, we obtain
\begin{equation}
  \lim_{\Lambda \to 0+} W_\Lambda \left[ \sqrt{R_\Lambda}\, \sigma
  \right]
  = \mathcal{W} [0] = - \ev \, \delta (0)\,.
\end{equation}
Thus, Eq.~(\ref{Appendix-WL}) gives the desired limit
\begin{equation}
  \lim_{\Lambda \to 0+} S_\Lambda [\sigma]
  = - \ev\, \delta (0) - \frac{1}{2} \int_p \sigma (p) \sigma (-p)\,.
\end{equation}

\section{Brief review of the large $N$ limit} \label{App:review_large_N}

In this appendix we briefly review the large $N$ approximation in the
ERG by following the method adopted in \cite{Sonoda:2023ohb}, which is
based on the method introduced in \cite{Morris:1997xj} for the
effective potentials.  The large $N$ limit of the ERG has been
discussed by several authors; we refer the interested reader to
\cite{DAttanasio:1997yph,Morris:1997xj,Blaizot:2005xy,Litim:2018pxe}
and references therein for a complete list of references that 
certainly includes pioneering works such as \cite{Wilson:1972cf} and \cite{Wegner:1972ih}.

The main aim of this appendix is to derive
Eqs.~(\ref{cL-diffeq-largeN}-\ref{eq:flow_m2L_large_N}) in the main text.  We work in $D$
dimensions with $2<D<4$.  To obtain the large $N$ limit it is
convenient to adopt the 1PI formalism.  Let us consider the ansatz
\begin{eqnarray}
  \Gamma_{\Lambda} & = &
                         -\frac{1}{2}\int_p \phi^i (-p) p^2 \phi^i (p)
                         + N\Gamma_{I\Lambda}\left[\frac{\phi^{i}\phi^{i}}{2N}\right]\,. 
                         \label{eq:ansatz_large_N}
\end{eqnarray}
This is consistent at leading order in large $N$. We next introduce
the variable $\varphi$ defined by
$\varphi\equiv\frac{\phi^{i}\phi^{i}}{2N}$. In the large $N$ limit,
the ERG equation for $\Gamma_{I\Lambda}$ reads
\begin{eqnarray}
-\Lambda\partial_{\Lambda}\Gamma_{I\Lambda} & = & \frac{1}{2}
\int_p  \Lambda\partial_{\Lambda}R_{\Lambda} \left(p\right)
{\cal G}_{\Lambda;p,-p}\left[\varphi \right]\,,\label{eq:ERG_for_Gamma_I_large_N}
\end{eqnarray}
where
\begin{eqnarray}
\int_q {\cal G}_{\Lambda;p,-q}\left[\varphi \right]
\Biggr\lbrace \left(q^2+R_\Lambda \left(q\right)\right)
\delta\left(q-r\right)
-\frac{\delta \Gamma_{I\Lambda}\left[\varphi\right]}{\delta \varphi \left(q-r\right)}\Biggr\rbrace
=
\delta\left(p-r\right)\,.
\end{eqnarray}

It turns out simpler to work with the Legendre transform of
$\Gamma_{I\Lambda}$.  We introduce the functional
\begin{eqnarray}
  F_{\Lambda}\left[\sigma\right] & = & \Gamma_{I\Lambda}\left[\varphi\right]
                                       -\int_p \sigma \left(p\right)\varphi \left(-p\right)\,,
\end{eqnarray}
where $\sigma=\frac{\delta\Gamma_{I\Lambda}}{\delta\varphi}\left[\varphi\right]$.
It follows that
\begin{eqnarray}
\varphi\left(p\right) & = & -\frac{\delta F_{\Lambda}}{\delta\sigma \left(-p\right)}\,.
\end{eqnarray}
The ERG equation for $F_{\Lambda} [\sigma]$ is then given by
\begin{eqnarray}
-\Lambda\partial_{\Lambda}F_{\Lambda}\left[\sigma\right] & = & 
\frac{1}{2}\int_p 
\Lambda\partial_{\Lambda}R_{\Lambda} \left(p\right)
{\cal G}_{\Lambda;p,-p}\left[\sigma\right]\,,\label{eq:ERG_for_F-1}
\end{eqnarray}
where $\mathcal{G}_\Lambda$, regarded as a functional of $\sigma$,
satisfies
\begin{equation}
\int_q {\cal G}_{\Lambda;p,-q}\left[\sigma \right]
\Biggr\lbrace \left(q^2+R_\Lambda \left(q\right)\right)
\delta\left(q-r\right)
-\sigma (r-q)\Biggr\rbrace
=
\delta\left(p-r\right)\,,
\end{equation}
and can be expressed by a geometric series of $\sigma$.

A particular solution to (\ref{eq:ERG_for_F-1}) is given by
\begin{eqnarray}
  I_{\Lambda}\left[\sigma\right] & = &
                                       c_{\Lambda}\delta\left(0\right)+c_{1\Lambda}\sigma\left(0\right)\notag\\
 &  & +\sum_{i=2}^{\infty}\frac{1}{2n}\int_{p_{1}\cdots
      p_{n}}\sigma\left(p_{1}\right)\cdots\sigma\left(p_{n}\right)
      \delta\left(\sum_{i=1}^{n}p_{i}\right)I_{n\Lambda}\left(p_{1},\cdots,p_{n}\right)\,,
\end{eqnarray}
where
\begin{subequations}
\begin{align}
c_{\Lambda} &=  -\frac{1}{2} 
                  \int_{q}\log\left(\frac{q^{2}+R_{\Lambda}\left(q\right)}{q^{2}}
                  \right)\,,\\  
c_{1\Lambda} &= 
                   \frac{1}{2}\int_{q}
               \left(\frac{1}{q^{2}+R_{\Lambda}\left(q\right)}
               -\frac{1}{q^{2}}\right)\,,\\  
I_{n\Lambda} & =
               \int_{q}h_{\Lambda}\left(q\right)h_{\Lambda}\left(q+p_{1}\right)
               \cdots h_{\Lambda}\left(q+p_{1}+\cdots+p_{n-1}\right)\,,
\end{align}
\end{subequations}
with $h_{\Lambda}\left(q\right)\equiv1/\left(q^{2}+R_{\Lambda}\left(q\right)\right)$. 

The general solution to (\ref{eq:ERG_for_F-1}) is obtained as
\begin{eqnarray}
F_{\Lambda}\left[\sigma\right] & = &
                                     \tilde{F}\left[\sigma\right]+I_{\Lambda}\left[\sigma\right]\,, 
\end{eqnarray}
where $\tilde{F}\left[\sigma\right]$ is an arbitrary functional independent
of $\Lambda$. In the present case we consider
\begin{eqnarray}
\tilde{F}\left[\sigma\right] & \equiv &
                                        f_{0}\delta\left(0\right)+f_{1}\sigma\left(0\right)
                                        + \frac{1}{\lambda}
                                        \int_{p}\frac{1}{2}\sigma\left(p\right)\sigma\left(-p\right)\,, 
\end{eqnarray}
where $\lambda$ is a positive constant reminiscent of the
$\phi^{4}$-interaction coupling.  Finally, let us also write down in a
compact form the solution for the case of constant field $\sigma$.  We
find
\begin{eqnarray}
F_{\Lambda}\left(\sigma\right) & = &
\tilde{F}\left(\sigma\right) 
-\frac{1}{2}\int_{q}\Biggr[\log\frac{q^{2}-\sigma+R_{\Lambda}}{q^{2}}
                                     + \frac{\sigma}{q^{2}}\Biggr]\,,
\end{eqnarray}
where
\begin{equation}
  F_\Lambda [\sigma] = F_\Lambda (\sigma)\,\delta (0)\,,\quad
  \tilde{F} [\sigma] = \tilde{F} (\sigma)\, \delta (0)\,.
\end{equation}

The main aim of this appendix has been to derive
Eqs.~(\ref{cL-diffeq-largeN}-\ref{eq:flow_m2L_large_N}) in the main
text.  Let us work in the symmetric phase, where
$c_\Lambda = G_\Lambda \left(0\right)$ (see
Eq.~(\ref{eq:cLambda_via_G_Lambda_vL})).  According to the relations
among the various functionals detailed in section
\ref{sec:effective_potential}, we can write the RHS of
(\ref{cL-diffeq-largeN}) by employing the following equation:
\begin{eqnarray}
1+c_{2\Lambda} \left(p \right) = R_\Lambda\left(p
  \right) \left(p^2-\frac{\delta\Gamma_{I\Lambda}}{\delta \varphi}
  +R_\Lambda\left(p\right) \right)^{-1} \Bigr|_{\varphi=0}\,,
\end{eqnarray}
which implies Eq.~(\ref{CL-largeN}) after we identify $m_{\Lambda}^{2}$
with
\begin{equation}
  -\sigma_{{\rm
    os}}=-\frac{\delta\Gamma_{I\Lambda}}{\delta\varphi}\left[\varphi=0\right]
= m_{\Lambda}^{2}\,.
\end{equation}
The ERG equation associated with $\sigma_{\rm{os}}$ reads
\begin{equation}
-\Lambda\frac{d}{d\Lambda}\sigma_{{\rm os}}  =  \frac{1}{2} \left(
  -\partial_{\sigma}^{2}F_{\Lambda}\right)^{-1}
\int_{q}\frac{\Lambda\partial_{\Lambda}R_{\Lambda}}{\left(q^{2} -
    \sigma_{{\rm os}}+R_{\Lambda}\right)^{2}}\,,
\end{equation}
which reproduces Eq.~(\ref{eq:flow_m2L_large_N}) once expressed in
terms of $m_{\Lambda}^{2}$.

\begin{acknowledgments}
  C.~P.~thanks Kobe University for hospitality where this project was
  initiated and pursued.
\end{acknowledgments}

\bibliography{paper-constant}

\begin{thebibliography}{18}%
\makeatletter
\providecommand \@ifxundefined [1]{%
 \@ifx{#1\undefined}
}%
\providecommand \@ifnum [1]{%
 \ifnum #1\expandafter \@firstoftwo
 \else \expandafter \@secondoftwo
 \fi
}%
\providecommand \@ifx [1]{%
 \ifx #1\expandafter \@firstoftwo
 \else \expandafter \@secondoftwo
 \fi
}%
\providecommand \natexlab [1]{#1}%
\providecommand \enquote  [1]{``#1''}%
\providecommand \bibnamefont  [1]{#1}%
\providecommand \bibfnamefont [1]{#1}%
\providecommand \citenamefont [1]{#1}%
\providecommand \href@noop [0]{\@secondoftwo}%
\providecommand \href [0]{\begingroup \@sanitize@url \@href}%
\providecommand \@href[1]{\@@startlink{#1}\@@href}%
\providecommand \@@href[1]{\endgroup#1\@@endlink}%
\providecommand \@sanitize@url [0]{\catcode `\\12\catcode `\$12\catcode
  `\&12\catcode `\#12\catcode `\^12\catcode `\_12\catcode `\%12\relax}%
\providecommand \@@startlink[1]{}%
\providecommand \@@endlink[0]{}%
\providecommand \url  [0]{\begingroup\@sanitize@url \@url }%
\providecommand \@url [1]{\endgroup\@href {#1}{\urlprefix }}%
\providecommand \urlprefix  [0]{URL }%
\providecommand \Eprint [0]{\href }%
\providecommand \doibase [0]{http://dx.doi.org/}%
\providecommand \selectlanguage [0]{\@gobble}%
\providecommand \bibinfo  [0]{\@secondoftwo}%
\providecommand \bibfield  [0]{\@secondoftwo}%
\providecommand \translation [1]{[#1]}%
\providecommand \BibitemOpen [0]{}%
\providecommand \bibitemStop [0]{}%
\providecommand \bibitemNoStop [0]{.\EOS\space}%
\providecommand \EOS [0]{\spacefactor3000\relax}%
\providecommand \BibitemShut  [1]{\csname bibitem#1\endcsname}%
\let\auto@bib@innerbib\@empty
\bibitem [{\citenamefont {Wilson}\ and\ \citenamefont
  {Kogut}(1974)}]{Wilson:1973jj}%
  \BibitemOpen
  \bibfield  {author} {\bibinfo {author} {\bibfnamefont {K.~G.}\ \bibnamefont
  {Wilson}}\ and\ \bibinfo {author} {\bibfnamefont {J.~B.}\ \bibnamefont
  {Kogut}},\ }\bibfield  {title} {\enquote {\bibinfo {title} {{The
  Renormalization group and the epsilon expansion}},}\ }\href@noop {}
  {\bibfield  {journal} {\bibinfo  {journal} {Phys.~Rept.}\ }\textbf {\bibinfo
  {volume} {12}},\ \bibinfo {pages} {75--200} (\bibinfo {year}
  {1974})}\BibitemShut {NoStop}%
\bibitem [{\citenamefont {Wegner}\ and\ \citenamefont
  {Houghton}(1973)}]{Wegner:1972ih}%
  \BibitemOpen
  \bibfield  {author} {\bibinfo {author} {\bibfnamefont {Franz~J.}\
  \bibnamefont {Wegner}}\ and\ \bibinfo {author} {\bibfnamefont {Anthony}\
  \bibnamefont {Houghton}},\ }\bibfield  {title} {\enquote {\bibinfo {title}
  {{Renormalization group equation for critical phenomena}},}\ }\href {\doibase
  10.1103/PhysRevA.8.401} {\bibfield  {journal} {\bibinfo  {journal} {Phys.
  Rev. A}\ }\textbf {\bibinfo {volume} {8}},\ \bibinfo {pages} {401--412}
  (\bibinfo {year} {1973})}\BibitemShut {NoStop}%
\bibitem [{\citenamefont {Polchinski}(1984)}]{Polchinski:1983gv}%
  \BibitemOpen
  \bibfield  {author} {\bibinfo {author} {\bibfnamefont {J.}~\bibnamefont
  {Polchinski}},\ }\bibfield  {title} {\enquote {\bibinfo {title}
  {{Renormalization and Effective Lagrangians}},}\ }\href {\doibase
  10.1016/0550-3213(84)90287-6} {\bibfield  {journal} {\bibinfo  {journal}
  {Nucl. Phys.}\ }\textbf {\bibinfo {volume} {B231}},\ \bibinfo {pages}
  {269--295} (\bibinfo {year} {1984})}\BibitemShut {NoStop}%
\bibitem [{\citenamefont {Wetterich}(1993)}]{Wetterich:1992yh}%
  \BibitemOpen
  \bibfield  {author} {\bibinfo {author} {\bibfnamefont {C.}~\bibnamefont
  {Wetterich}},\ }\bibfield  {title} {\enquote {\bibinfo {title} {{Exact
  evolution equation for the effective potential}},}\ }\href {\doibase
  10.1016/0370-2693(93)90726-X} {\bibfield  {journal} {\bibinfo  {journal}
  {Phys. Lett.}\ }\textbf {\bibinfo {volume} {B301}},\ \bibinfo {pages}
  {90--94} (\bibinfo {year} {1993})}\BibitemShut {NoStop}%
\bibitem [{\citenamefont {Reuter}\ and\ \citenamefont
  {Wetterich}(1994)}]{Reuter:1993kw}%
  \BibitemOpen
  \bibfield  {author} {\bibinfo {author} {\bibfnamefont {M.}~\bibnamefont
  {Reuter}}\ and\ \bibinfo {author} {\bibfnamefont {C.}~\bibnamefont
  {Wetterich}},\ }\bibfield  {title} {\enquote {\bibinfo {title} {{Effective
  average action for gauge theories and exact evolution equations}},}\ }\href
  {\doibase 10.1016/0550-3213(94)90543-6} {\bibfield  {journal} {\bibinfo
  {journal} {Nucl. Phys.}\ }\textbf {\bibinfo {volume} {B417}},\ \bibinfo
  {pages} {181--214} (\bibinfo {year} {1994})}\BibitemShut {NoStop}%
\bibitem [{\citenamefont {Morris}(1994)}]{Morris:1993qb}%
  \BibitemOpen
  \bibfield  {author} {\bibinfo {author} {\bibfnamefont {T.~R.}\ \bibnamefont
  {Morris}},\ }\bibfield  {title} {\enquote {\bibinfo {title} {{The Exact
  renormalization group and approximate solutions}},}\ }\href {\doibase
  10.1142/S0217751X94000972} {\bibfield  {journal} {\bibinfo  {journal} {Int.
  J. Mod. Phys.}\ }\textbf {\bibinfo {volume} {A9}},\ \bibinfo {pages}
  {2411--2450} (\bibinfo {year} {1994})},\ \Eprint
  {http://arxiv.org/abs/hep-ph/9308265} {arXiv:hep-ph/9308265 [hep-ph]}
  \BibitemShut {NoStop}%
\bibitem [{\citenamefont {Wegner}(1974)}]{Wegner_1974}%
  \BibitemOpen
  \bibfield  {author} {\bibinfo {author} {\bibfnamefont {F.~J.}\ \bibnamefont
  {Wegner}},\ }\bibfield  {title} {\enquote {\bibinfo {title} {Some invariance
  properties of the renormalization group},}\ }\href {\doibase
  10.1088/0022-3719/7/12/004} {\bibfield  {journal} {\bibinfo  {journal}
  {Journal of Physics C: Solid State Physics}\ }\textbf {\bibinfo {volume}
  {7}},\ \bibinfo {pages} {2098} (\bibinfo {year} {1974})}\BibitemShut
  {NoStop}%
\bibitem [{\citenamefont {Sonoda}(2015)}]{Sonoda:2015bla}%
  \BibitemOpen
  \bibfield  {author} {\bibinfo {author} {\bibfnamefont {H.}~\bibnamefont
  {Sonoda}},\ }\bibfield  {title} {\enquote {\bibinfo {title} {{Equivalence of
  Wilson Actions}},}\ }\href {\doibase 10.1093/ptep/ptv130} {\bibfield
  {journal} {\bibinfo  {journal} {PTEP}\ }\textbf {\bibinfo {volume} {2015}},\
  \bibinfo {pages} {103B01} (\bibinfo {year} {2015})},\ \Eprint
  {http://arxiv.org/abs/1503.08578} {arXiv:1503.08578 [hep-th]} \BibitemShut
  {NoStop}%
\bibitem [{\citenamefont {Reuter}(1998)}]{Reuter:1996cp}%
  \BibitemOpen
  \bibfield  {author} {\bibinfo {author} {\bibfnamefont {M.}~\bibnamefont
  {Reuter}},\ }\bibfield  {title} {\enquote {\bibinfo {title} {{Nonperturbative
  evolution equation for quantum gravity}},}\ }\href {\doibase
  10.1103/PhysRevD.57.971} {\bibfield  {journal} {\bibinfo  {journal} {Phys.
  Rev.}\ }\textbf {\bibinfo {volume} {D57}},\ \bibinfo {pages} {971--985}
  (\bibinfo {year} {1998})},\ \Eprint {http://arxiv.org/abs/hep-th/9605030}
  {arXiv:hep-th/9605030 [hep-th]} \BibitemShut {NoStop}%
\bibitem [{\citenamefont {Reuter}\ and\ \citenamefont
  {Saueressig}(2019)}]{Reuter_Saueressig_2019}%
  \BibitemOpen
  \bibfield  {author} {\bibinfo {author} {\bibfnamefont {Martin}\ \bibnamefont
  {Reuter}}\ and\ \bibinfo {author} {\bibfnamefont {Frank}\ \bibnamefont
  {Saueressig}},\ }\href@noop {} {\emph {\bibinfo {title} {Quantum Gravity and
  the Functional Renormalization Group: The Road towards Asymptotic Safety}}},\
  Cambridge Monographs on Mathematical Physics\ (\bibinfo  {publisher}
  {Cambridge University Press},\ \bibinfo {year} {2019})\BibitemShut {NoStop}%
\bibitem [{\citenamefont {Percacci}(2017)}]{Percacci_book_2017}%
  \BibitemOpen
  \bibfield  {author} {\bibinfo {author} {\bibfnamefont {Roberto}\ \bibnamefont
  {Percacci}},\ }\href {\doibase 10.1142/10369} {\emph {\bibinfo {title} {An
  Introduction to Covariant Quantum Gravity and Asymptotic Safety}}}\ (\bibinfo
   {publisher} {World Scientific},\ \bibinfo {year} {2017})\ \Eprint
  {http://arxiv.org/abs/https://www.worldscientific.com/doi/pdf/10.1142/10369}
  {https://www.worldscientific.com/doi/pdf/10.1142/10369} \BibitemShut
  {NoStop}%
\bibitem [{\citenamefont {Sonoda}(2017)}]{Sonoda:2017rro}%
  \BibitemOpen
  \bibfield  {author} {\bibinfo {author} {\bibfnamefont {H.}~\bibnamefont
  {Sonoda}},\ }\bibfield  {title} {\enquote {\bibinfo {title} {{The generating
  functional of correlation functions as a high momentum limit of a Wilson
  action}},}\ }\href {\doibase 10.1093/ptep/ptx152} {\bibfield  {journal}
  {\bibinfo  {journal} {PTEP}\ }\textbf {\bibinfo {volume} {2017}},\ \bibinfo
  {pages} {123B01} (\bibinfo {year} {2017})},\ \Eprint
  {http://arxiv.org/abs/1706.00198} {arXiv:1706.00198 [hep-th]} \BibitemShut
  {NoStop}%
\bibitem [{\citenamefont {Sonoda}(2023)}]{Sonoda:2023ohb}%
  \BibitemOpen
  \bibfield  {author} {\bibinfo {author} {\bibfnamefont {Hidenori}\
  \bibnamefont {Sonoda}},\ }\bibfield  {title} {\enquote {\bibinfo {title}
  {{Exact Renormalization Group in Large $N$}},}\ }\href@noop {} {\  (\bibinfo
  {year} {2023})},\ \Eprint {http://arxiv.org/abs/2302.09914} {arXiv:2302.09914
  [hep-th]} \BibitemShut {NoStop}%
\bibitem [{\citenamefont {Morris}\ and\ \citenamefont
  {Turner}(1998)}]{Morris:1997xj}%
  \BibitemOpen
  \bibfield  {author} {\bibinfo {author} {\bibfnamefont {Tim~R.}\ \bibnamefont
  {Morris}}\ and\ \bibinfo {author} {\bibfnamefont {Michael~D.}\ \bibnamefont
  {Turner}},\ }\bibfield  {title} {\enquote {\bibinfo {title} {{Derivative
  expansion of the renormalization group in O(N) scalar field theory}},}\
  }\href {\doibase 10.1016/S0550-3213(97)00640-8} {\bibfield  {journal}
  {\bibinfo  {journal} {Nucl. Phys. B}\ }\textbf {\bibinfo {volume} {509}},\
  \bibinfo {pages} {637--661} (\bibinfo {year} {1998})},\ \Eprint
  {http://arxiv.org/abs/hep-th/9704202} {arXiv:hep-th/9704202} \BibitemShut
  {NoStop}%
\bibitem [{\citenamefont {D'Attanasio}\ and\ \citenamefont
  {Morris}(1997)}]{DAttanasio:1997yph}%
  \BibitemOpen
  \bibfield  {author} {\bibinfo {author} {\bibfnamefont {Marco}\ \bibnamefont
  {D'Attanasio}}\ and\ \bibinfo {author} {\bibfnamefont {Tim~R.}\ \bibnamefont
  {Morris}},\ }\bibfield  {title} {\enquote {\bibinfo {title} {{Large N and the
  renormalization group}},}\ }\href {\doibase 10.1016/S0370-2693(97)00866-6}
  {\bibfield  {journal} {\bibinfo  {journal} {Phys. Lett. B}\ }\textbf
  {\bibinfo {volume} {409}},\ \bibinfo {pages} {363--370} (\bibinfo {year}
  {1997})},\ \Eprint {http://arxiv.org/abs/hep-th/9704094}
  {arXiv:hep-th/9704094} \BibitemShut {NoStop}%
\bibitem [{\citenamefont {Blaizot}\ \emph {et~al.}(2006)\citenamefont
  {Blaizot}, \citenamefont {Mendez~Galain},\ and\ \citenamefont
  {Wschebor}}]{Blaizot:2005xy}%
  \BibitemOpen
  \bibfield  {author} {\bibinfo {author} {\bibfnamefont {J.~P.}\ \bibnamefont
  {Blaizot}}, \bibinfo {author} {\bibfnamefont {Ramon}\ \bibnamefont
  {Mendez~Galain}}, \ and\ \bibinfo {author} {\bibfnamefont {Nicolas}\
  \bibnamefont {Wschebor}},\ }\bibfield  {title} {\enquote {\bibinfo {title}
  {{A New method to solve the non perturbative renormalization group
  equations}},}\ }\href {\doibase 10.1016/j.physletb.2005.10.086} {\bibfield
  {journal} {\bibinfo  {journal} {Phys. Lett. B}\ }\textbf {\bibinfo {volume}
  {632}},\ \bibinfo {pages} {571--578} (\bibinfo {year} {2006})},\ \Eprint
  {http://arxiv.org/abs/hep-th/0503103} {arXiv:hep-th/0503103} \BibitemShut
  {NoStop}%
\bibitem [{\citenamefont {Litim}\ and\ \citenamefont
  {Trott}(2018)}]{Litim:2018pxe}%
  \BibitemOpen
  \bibfield  {author} {\bibinfo {author} {\bibfnamefont {Daniel~F.}\
  \bibnamefont {Litim}}\ and\ \bibinfo {author} {\bibfnamefont {Matthew~J.}\
  \bibnamefont {Trott}},\ }\bibfield  {title} {\enquote {\bibinfo {title}
  {{Asymptotic safety of scalar field theories}},}\ }\href {\doibase
  10.1103/PhysRevD.98.125006} {\bibfield  {journal} {\bibinfo  {journal} {Phys.
  Rev. D}\ }\textbf {\bibinfo {volume} {98}},\ \bibinfo {pages} {125006}
  (\bibinfo {year} {2018})},\ \Eprint {http://arxiv.org/abs/1810.01678}
  {arXiv:1810.01678 [hep-th]} \BibitemShut {NoStop}%
\bibitem [{\citenamefont {Wilson}(1973)}]{Wilson:1972cf}%
  \BibitemOpen
  \bibfield  {author} {\bibinfo {author} {\bibfnamefont {Kenneth~G.}\
  \bibnamefont {Wilson}},\ }\bibfield  {title} {\enquote {\bibinfo {title}
  {{Quantum field theory models in less than four-dimensions}},}\ }\href
  {\doibase 10.1103/PhysRevD.7.2911} {\bibfield  {journal} {\bibinfo  {journal}
  {Phys. Rev. D}\ }\textbf {\bibinfo {volume} {7}},\ \bibinfo {pages}
  {2911--2926} (\bibinfo {year} {1973})}\BibitemShut {NoStop}%
\end{thebibliography}%

\end{document}